\documentclass[aps,twocolumn,showpacs,amsmath,amssymb,superscriptaddress,prl]{revtex4-1}
\usepackage{graphicx}
\usepackage{dcolumn}
\usepackage{bm}
\usepackage{siunitx}

\providecommand{\e}[1]{\ensuremath{\times 10^{#1}}}

\begin{document}

\title{Nonextensive statistics in spin precession}

\author{M.J.~Bales}
\email[Corresponding author: ]{matthew.bales@tum.de}
\affiliation{Physikdepartment, Technische Universit{\"a}t M{\"u}nchen, D-85748 Garching, Germany}
\author{P. Fierlinger}
\affiliation{Physikdepartment, Technische Universit{\"a}t M{\"u}nchen, D-85748 Garching, Germany}
\author{R. Golub}
\affiliation{Physics Department, North Carolina State University, Raleigh, NC 27695-8202, USA}

\date{\today}


\begin{abstract}
Many experiments utilize the precession of trapped particles in magnetic fields to perform high precision measurements. It had been presumed that after free precession, initially polarized particles will form a Gaussian phase distribution in the plane of precession. We show that trapped particles in the presence of magnetic field gradients and electric fields will often form a non-Gaussian distribution with power-law tails which are consistent with nonextensive statistics. As the exact shape of the distribution depends upon many experimental parameters, it provides a potential new technique to directly measure them.
\end{abstract}

\maketitle{}

	The Larmor spin precession of trapped particles is used in many experimental systems such as spin clocks, magnetometry~\cite{Budker2007}, and for the detection of electric dipole moments (EDMs) in elementary particles~\cite{nEDMNewLimit, nEDMFRMII, nEDMNewLimit}.   In particular, EDM experiments in atomic and particle systems require precise understanding of precession~\cite{Pendlebury2004, Lamoreaux2005, Swank, Pignol2012, Steyerl1,PignolGeneral,Golub2015, Golub2014Relax} to explore beyond the Standard Model parameters at energy ranges greater than one TeV~\cite{EDMGlobal,LeptonDipoleMomentsBook}.   While it is well known that magnetic field gradients and electric fields create relaxation and frequency shifts, it had been assumed that after free precession~\cite{Lamoreaux2005, Barabanov, SpinRelaxE}, initially polarized particles will form a Gaussian phase distribution in the plane of precession.  In this study, we describe a previously unidentified phenomenon which causes non-Gaussian phase distributions to form.  Simulations of spin precession resulted in phase distributions which fit well to distributions predicted by nonextensive statistics~\cite{nonextensive1, nonextensive2, DeVoe2009, Douglas2006}, where noise is not simply additive, but instead partially multiplicative.  The shape of these phase distributions depends on a rich interplay of many experimental parameters and could potentially be used to quantify them.

	The precession of the dipole of an individual particle will evolve classically between collisions with the chamber,
	
\begin{equation}
\label{eq:blochphi}
\frac{d\phi}{dt}=\gamma \left( \left(B_x(t) \cos\phi + B_y(t) \sin\phi\right)\tan\theta -B_z(t)\right)
\end{equation}

\begin{equation}
\label{eq:blochtheta}
\frac{d\theta}{dt}=\gamma  \left(B_y(t)  \cos\phi - B_x(t) \sin\phi\right)
\end{equation}

	where $\phi$ is the angle of the particle's spin in the plane of precession, $\theta$ is the angle of the particle's spin with respect to the precession plane (latitude), $\gamma$ is the gyromagnetic ratio of the particle, and $\mathbf{B}(t)$ is the magnetic field as seen by the particle~\cite{Pendlebury2004}. Here any EDM is presumed to be small. In an ideal spin precession experiment, a constant magnetic field $\mathbf{B_0}$ would be applied in the $z$ direction, perpendicular to the precession plane, so that $d\phi/dt=\omega_0$ where $\omega_0 \equiv \gamma |\mathbf{B_0}|$ is the Larmor frequency.  +
	
	Inevitably, transverse magnetic fields $\mathbf{B_{xy}}$ are present in an experiment. When $\mathbf{B_{xy}}$ is constant across the chamber, only the rotation axis would be altered.  When $\mathbf{B_{xy}}$ is time dependent in the particle's rest frame, then Eq.~\ref{eq:blochphi} demonstrates that particles will precess at different rates if $\theta\neq0$ and Eq.~\ref{eq:blochtheta} shows that their spin will leave the precession plane.
		 \begin{figure}
 \includegraphics[width=\linewidth]{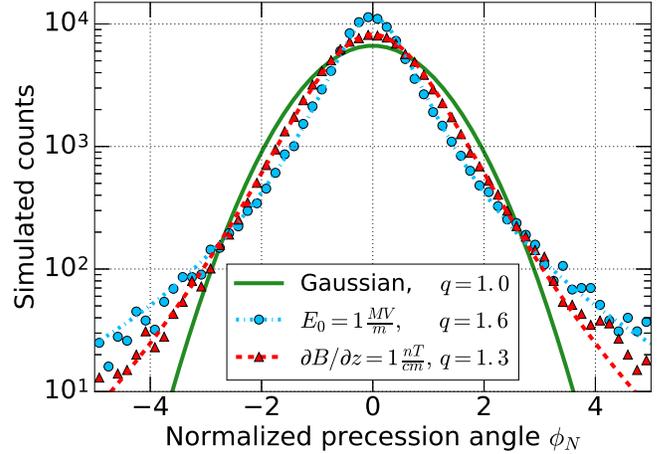}
 \caption
 	{\label{fig:Distributions}
	Two simulated spin precession phase histograms for the $\phi$ phase angle fitted by Tsallis q-Gaussian distributions are plotted for $^{199}$Hg in the presence of either an electric field (blue) or a linear magnetic field gradient (red).  The histograms are binned by their normalized precession phase $\phi_N$ such that $\phi_N=0$ is the mean precession phase and $\phi_N=1$ represents one standard deviation $\sigma_\phi$.  The 3D simulations precessed 100,000 $^{199}$Hg atoms for 100 s.  For $E_0$=\SI[per-mode=symbol]{1}{\mega\volt\per\meter}: $\sigma_\phi$=3.9\e{-6} radians and   $q$=1.6. For $\partial B / \partial z$=\SI[per-mode=symbol]{1}{\nano\tesla\per\centi\meter}: $\sigma_\phi$=5.7\e{-2} radians and $q$=1.3  A Gaussian distribution (green) is shown for comparison.
 	}
 \end{figure}
 
 \begin{figure*}
 \includegraphics[width=\textwidth]{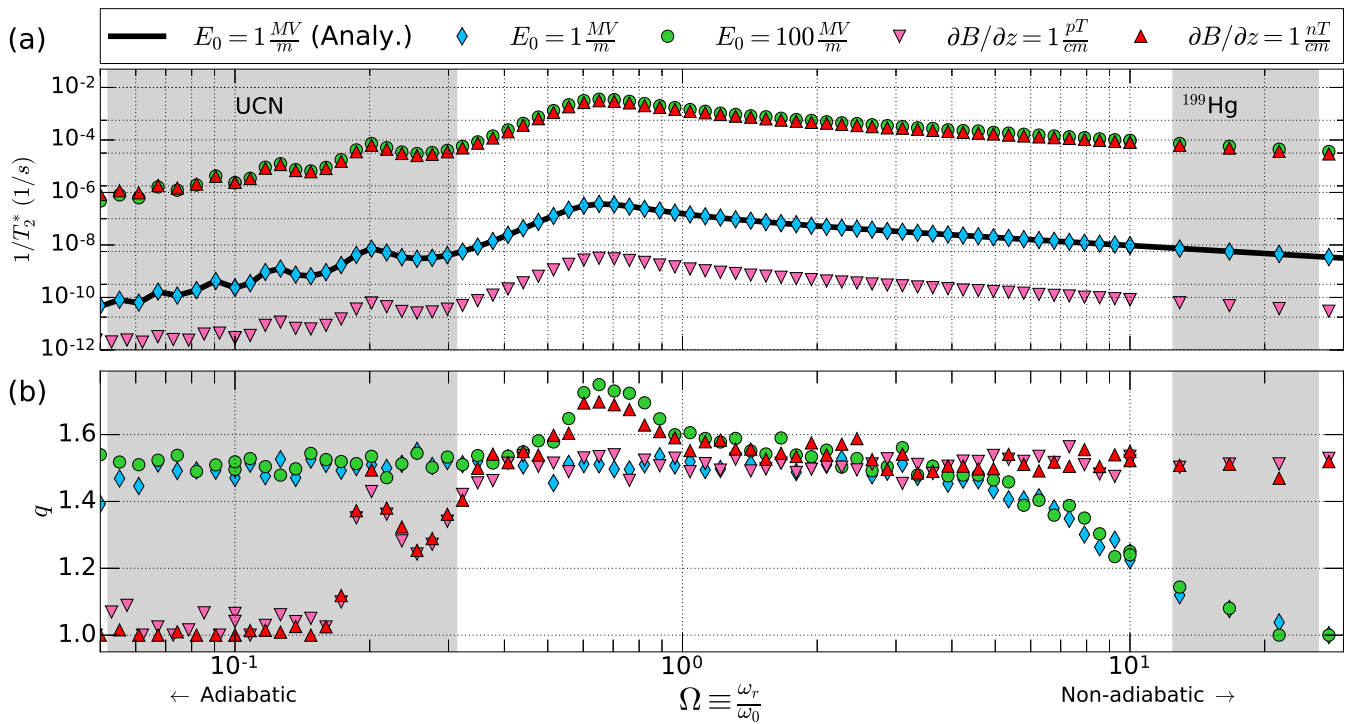}
 \caption
 	{\label{fig:T2}
	Plotted are (a) the inverse $T_2^*$ time contribution from the motional magnetic field and linear magnetic field gradients and (b) the fitted q-Gaussian power $q$ of the $\phi$ distribution versus the ratio $\Omega$ of orbital frequency and the precession frequency, which was varied by altering the simulated velocity.  Shown in grey are the approximate regions where the majority of $^{199}$Hg atoms and UCNs are found in an EDM experiment when the magnetic field strength is \SI[per-mode=symbol]{1}{\micro\tesla}. Each point represents the result of a 2D simulation which precessed 10,000 $^{199}$Hg atoms for 100 s.  Also shown (black line) is an analytical calculation using the method of~\cite{Golub2015, Golub2014Relax}.
 	}
 \end{figure*}
	Magnetic field gradients and electric fields $\mathbf{E}$ both cause particles to experience a time dependent transverse magnetic field $\mathbf{B_{xy}}(t)$.  For electric fields, $\mathbf{B_{xy}}(t)$ is created by a motional magnetic field  $\mathbf{B_v}(t) \propto \mathbf{E} \times \mathbf{v}(t)$ where $\mathbf{v}(t)$ is the particle's velocity. Either source of $\mathbf{B_{xy}}(t)$ causes free induction decay (FID) of the measured signal. This is the result of spin relaxation~\cite{Golub2014Relax, Barabanov, SpinRelaxE}, which is the dephasing of the spin ensemble over time and appears as noise in a frequency measurement.  This is one of the reasons significant efforts are made in spin precession experiments to minimize magnetic field gradients~\cite{shieldedroom, MagMin, insert}. Spin relaxation can also be caused by interactions with other particles or with the chamber walls, but we restrict our discussion to systems which are dominated by relaxation caused by $\mathbf{B_{xy}}(t)$. When the particle reflects diffusely from the chamber walls, the random motion of the particles trajectory over the precession period $\tau$ results in an exponential depolarization often parameterized by a decay time constant $T_2^*$. It has generally been presumed that phase distributions produced after free precession were Gaussian in nature, except under the influence of a significant asymmetry such as gravity~\cite{GravDepol1}.
		
	Classically, the non-Gaussian distributions observed in the simulations (see Fig~\ref{fig:Distributions}) originate from a cascading effect from the spins' motion out of the precession plane, $|\theta| > 0$.  From Eq.~\ref{eq:blochphi}, it is apparent that subsequent precession in the $\phi$ direction is enhanced by a factor of $\tan \theta$.   Particles which precess out of the plane will experience an accordingly larger phase shift for a given trajectory through the transverse magnetic fields, enhancing the relaxation for these particles.  So while a Gaussian distribution will form in $\theta$, a non-Gaussian distribution will form in $\phi$. This will also result in a correlation between the $\theta$ and the $\phi$ distributions resulting from precession, which must be accounted for when determining the effect on a frequency measurement. When spin precession in the $\theta$ direction was completely suppressed, the simulations produced Gaussian $\phi$ distributions with widths proportional to the $\mathbf{B_{xy}}(t)$ fields, as naively expected from the Central Limit theorem.  The source of the multiplicative noise can also be seen by looking at the second order solution of the Schroedinger equation, which yields, for arbitrary time dependence of the transverse fields, the frequency shift~\cite{Golub2014Relax,Golub2015}
		
\begin{equation}
\delta\omega=\frac{d\phi}{dt}=2\operatorname{Im}\int_{0}^{t}d\tau 
e^{i\omega_{0}\tau}\left(  \bar{\omega}^{\ast}\left(  t\right)  \bar{\omega}\left(
t-\tau\right)  \right)
\end{equation}

where $\bar{\omega}\left(  t\right)  = \gamma \left(B_{x}\left(  t\right)  +i B
_{y}\left(  t\right)\right)  $ is, in general, a stochastic variable. This is an example of a generalized Langevin equation~\citep{Uneyama2015}. For more information see also Redner~\cite{Redner1990}.

The simulations that revealed the general non-Gaussian behavior were performed with our own custom built C++ software which utilized both \textsc{ROOT}~\cite{ROOT} and \textsc{BOOST ODEINT}~\cite{BOOSTODEINT} libraries. Particles in the simulation were first tracked in ballistic trajectories inside a 2D or 3D cylindrical chamber. Wall reflections were modeled with a tunable diffusivity $D$ where $D$=1 was 100\% diffuse Lambert reflections, which are distributed by $\cos\theta_r$ where $\theta_r$ is the reflected angle and are independent of the angle of incidence,  and $D$=0 was purely specular. The tracks through the chamber were then used to precess the particles' spins by integrating the equations of motion, Eq.~\ref{eq:blochphi} and Eq.~\ref{eq:blochtheta}, using a dynamic Runge-Kutta algorithm. Inter-particle collisions, absorption/depolarization from chamber wall interactions, and gravity were ignored. Simulated results were consistent with comparisons to a constant stepper algorithm, a 128-bit floating point method, and our own \textsc{Geant4}~\cite{geant4} simulation of spin precession.  

The software was designed to study neutron EDM experiments~\cite{nEDMNewLimit, nEDMFRMII, nEDMNewLimit}  which utilize ultracold neutrons (UCNs) and room temperature comagnetometers (e.g. $^{199}$Hg) in the ballistic regime where inter-particle interactions are minimal.  The software was validated extensively against analytical predictions of frequency shifts and relaxation, including predictions of the geometric phase effect in EDM experiments~\cite{Berry2010, Pendlebury2004, Lamoreaux2005, Swank, Pignol2012, Steyerl1}.  It should be noted that the simulations were able to accurately produce phase shifts \ensuremath{< 10^{-6}} rad and decay rates \ensuremath{< 10^{-8}} \SI[]{}{\per\second} (See Fig.~\ref{fig:T2} (a)).  In the following discussions, the parameters used, unless otherwise indicated, were: the Sussex neutron EDM experiment's chamber dimensions (radius r = 0.235 m, height = 0.12 m)~\cite{nEDMNewLimit}, $D$=1, uniform fields, $B_0$=\SI[per-mode=symbol]{1}{\micro\tesla}, $\gamma$= \SI[]{-48.46}{ \per\micro\tesla \per \second} ($^{199}$Hg), and $\tau$ = 100 s.

The simulated non-Gaussian $\phi$ distributions were observed under a variety of simulated conditions providing sufficient free precession time was established (see Fig.~\ref{fig:Distributions}) At early precession times, the distribution caused by $\mathbf{B_{xy}}(t)$ fields appeared nearly Gaussian.  However after sufficient time the $\phi$ distribution could be well fit at its mean phase by a Tsallis q-Gaussian distribution~\cite{nonextensive1}

\begin{equation}
\label{eq:Tsallis}
F_T (x) = A_0 \left( 1 + \left( 1 - q \right) \beta x^2 \right)^{-1 / (1-q)}
\end{equation}

with free parameters $A_0$ , $\beta$, and $q$. As $q \to 1$, $F_T$ will approach a Gaussian distribution, $A_0 e^{-\beta x^2}$. However for $q > 1$ the distribution contains power law tails with a power of $2/(1 - q)$.  It should be noted that a q-Gaussian distribution convolved with additional Gaussian noise appears to produce a q-Gaussian distribution, but with a lower $q$.

Typically in a q-Gaussian distribution, the power $q$ represents the entropic index~\cite{nonextensive1}.  When $q=1$, the system is described by classical Boltzmann-Gibbs statistical mechanics.  When $q \neq 1$, the system is described by the more general nonextensive statistical mechanics.  In nonextensive statistics, entropy is not only additive but also multiplicative.  In this work, we report on the observed dependence of the fitted q-Gaussian power $q$ by varying parameters in our simulations. We cannot yet state an underlying analytical theory which predicts the value $q$ takes.

Our simulations in 2D showed resonant behavior between the particles' orbital frequency around the chamber $\omega_r \equiv v_{xy} / r$ and the Larmor frequency $\omega_0$.  When the ratio $\Omega \equiv \omega_r / \omega_0$ was approximately 0.64, there was a dramatic decrease in $T_2^*$ (see Fig.~\ref{fig:T2} (a)). The region where $\Omega \ll 1$ is referred to as the adiabatic regime, where UCN experiments typically operate.  The region where $\Omega \gg 1$ is the nonadiabatic regime, where experiments which utilize room temperature atoms and molecules, such as $^{199}$Hg, typically operate.  Resonant phenomena were clearly present in the adiabatic regime for both motional magnetic fields and magnetic gradients. The results of an analytic calculation for a motional  magnetic field using the method of  ~\cite{Golub2015, Golub2014Relax} show remarkable agreement with the simulation.  It should be noted that 3D simulations of Maxwellian velocity profiles differ significantly from 2D simulations due to the large range of particle velocities and the variation of magnetic field strengths in the $z$ direction for the vertical magnetic gradient. For example, the non-adiabatic regime exhibited roughly an order of magnitude decreased $T_2^*$.

Fits of q-Gaussian distributions to simulated $\phi$ distributions also showed distinct $\Omega$ dependent behavior in 2D  (see Fig.~\ref{fig:T2} (b)).  For weaker $\mathbf{B_{xy}}(t)$ fields, from either motional fields or magnetic field gradients, $q$ approached $\approx$1.5 in the region near $\Omega=1$.  The behavior differed between the two cases in the adiabatic and non-adiabatic regions.  For stronger $\mathbf{B_{xy}}(t)$ fields, a larger non-Gaussian behavior was seen at the same resonant $\Omega$ as seen in Fig.~\ref{fig:T2} (a). Simulations for shorter precession times caused the stable $q\approx1.5$ region to shrink for either case. 

The non-Gaussian behavior persists in cases more complex than the simple 2D case.  In 3D simulations with Maxwell velocity profiles, a small skewness (the third standardized moment of the distribution), was introduced to the $\phi$ distributions as particles with different velocities $v_{xy}$ perpendicular to the field experience different second order phase shifts from the $\mathbf{B_{xy}}(t)$ fields~\cite{Steyerl1}.  This skewness increased over time and is barely discernible in Fig.~\ref{fig:Distributions}.  An increased chance of specular reflections both increased the skewness of distributions and decreased the $q$ of simulated distributions.

\begin{figure}
 \includegraphics[width=\linewidth]{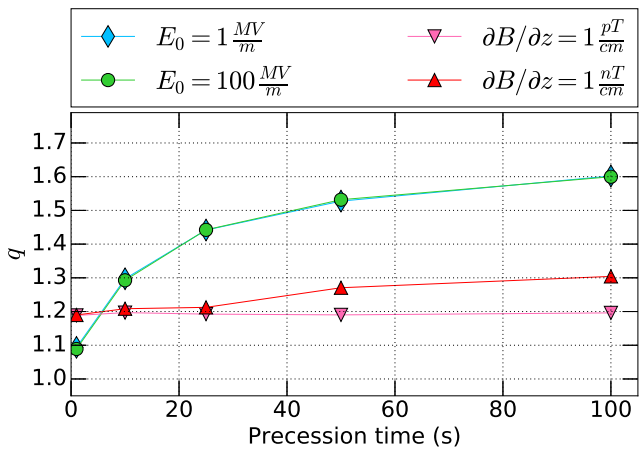}
 \caption
 	{ \label{fig:TimeEvolution}
	Time evolution of the fitted q-Gaussian power $q$ in either motional magnetic fields induced by an electric field or from magnetic field gradients. Each point represents the result of a 3D simulation which precessed 100,000 $^{199}$Hg atoms at room temperature, 293.15 K, for 100 s. The two electric field strengths overlap almost completely on the plot. Lines are drawn between data points to aid the eye.
 	}
 \end{figure}

For very short precession periods, particles do not uniformly sample the chamber, and their distributions therefore are not Tsallis q-Gaussian.  For $^{199}$Hg at room temperature in a 1~\si[per-mode=symbol]{\micro\tesla} $\mathbf{B_0}$ field, the q-Gaussian distribution resolved after approximately 0.1~\si[per-mode=symbol]{\second}.  After this time, the distribution's fitted $q$ evolved differently depending on the presence of a magnetic field gradient or an electric field (see Fig.~\ref{fig:TimeEvolution}).

\begin{figure}
 \includegraphics[width=\linewidth]{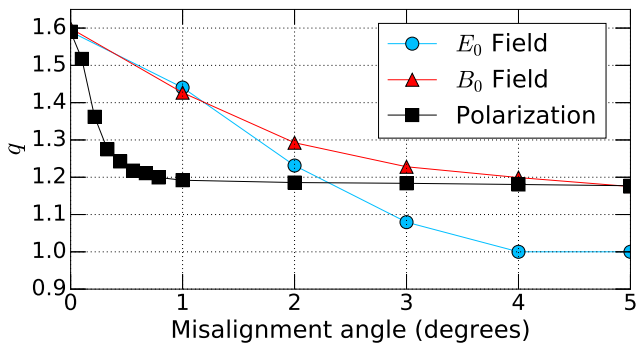}
 \caption
 	{ \label{fig:MisalignmentAngles}
	The effect of misalignment of the vectors which define the initial polarization, the $B_0$ field, and the $E_0$ field on the fitted q-Gaussian power $q$ from motional magnetic fields induced by an electric field.   Here ``0 degrees'' represents exactly parallel or anti-parallel fields and a perpendicular polarization.  Each point represents the result of a 3D simulation which precessed 100,000 $^{199}$Hg atoms at room temperature, 293.15 K, for 100 s. Lines are drawn between data points to aid the eye.
 	}
 \end{figure}

One possible suppression of the non-Gaussian behavior was found by altering the angles between the vectors defining $\mathbf{E_0}$ and $\mathbf{B_0}$, which should nominally be parallel or anti-parallel, and the initial polarization, which is nominally perpendicular to the other two.  Altering the alignment of these vectors individually, we see in Fig.~\ref{fig:MisalignmentAngles} that the distribution remains non-Gaussian for relatively large angles.  In the presence of a linear magnetic field gradient (not shown), misalignment angles up to 5 degrees were found to minimally impact the distribution's shape, which remained non-Gaussian.
 
To directly measure the non-Gaussian distributions in an experiment, a system with a long $T_2^*$ at low magnetic field gradients would be essential, such as $^{199}$Hg, UCNs, $^{3}$He, or $^{129}$Xe.  The measurement itself could take place in a constant vertical magnetic field gradient which is both strong enough to create significant relaxation, but not so large that the resulting phase distribution would span $2\pi$.  For $^{199}$Hg, a field on the order of \SI[per-mode=symbol]{1}{\nano \tesla \per \cm} could be ideal. The distribution could then be mapped out with repeated measurements after long precession times.  A continuous frequency measurement would also reveal this behavior under transformation to Fourier space, resulting in a modified Lorentzian distribution. 

There is tantalizing evidence of this effect in an existing measurement of the neutron EDM.  Fig 5. showing its most recent analysis~\cite{nEDMNewLimit} shows a plot of normalized frequency measurements that appears to hint at the presence of q-Gaussian distribution.  Given that the $q$ value is visibly small and that the authors comment that the data outside 4 $\sigma$ could be due to runs with known issues, these results are not conclusive evidence.  The simulations presented here do not include gravity or other details of this experiment, which could significantly impact the distribution's shape.

There is still more to explore in relation to phase distributions after precession. The effects of particle interactions, mean free path, and gravity on these distributions still need to be studied.   The distributions could possibly be used to measure experimental parameters as their behavior is sensitive to many relevant parameters such as initial polarization, particle velocity, field alignment angles, precession time, field strengths, and gradient strengths.  Further studies will be required to quantify these sensitivities. While this work has examined trapped particles confined within a chamber, nonextensive statistics arising from motion out of the precession plane could also appear in other spin systems such as experiments measuring the muon EDM, the proton EDM,  and the anomalous magnetic moment of the muon (g-2).  We have shown in this work, for the first time, that Larmor spin precession of trapped particles forms a non-Gaussian phase distribution which is consistent with nonextensive statistics and that there is a rich behavior in their phase distributions that must be explored. 

\begin{acknowledgments}
We thank Douglas Beck, Timothy Chupp, Skyler Degenkolb, and William Terrano for valuable discussions. This work was supported in part by the Cluster of Excellence ``Origin and Structure of the Universe'' at the Technische Universit\"{a}t M\"{u}nchen, the DFG Priority Program SPP 1491 ``Precision Experiments with Cold and Ultra-Cold Neutrons'', and  by the US Department of Energy under Grant No. DE-FG02-97ER41042.
\end{acknowledgments}

\end{document}